\documentclass[aps,prb,twocolumn,amsmath,superscriptaddress]{revtex4}
\usepackage{graphicx}
\usepackage{color}

\begin{document}

\title{Resonant optical pumping of a Mn spin in a strain free quantum dot}

\author{L. Besombes}
\email{lucien.besombes@grenoble.cnrs.fr}
\affiliation{CEA-CNRS group "Nanophysique et
semiconducteurs", CNRS, Institut N\'eel, F-38042 Grenoble,
France.}

\author{H. Boukari}
\affiliation{CEA-CNRS group "Nanophysique et
semiconducteurs", CNRS, Institut N\'eel, F-38042 Grenoble,
France.}

\date{\today}

\begin{abstract}

We report on the spin properties of individual Mn atoms in II-VI semiconductor strain free quantum dots. Strain free Mn-doped CdTe quantum dots are formed by width fluctuations in thin quantum wells lattice matched on a CdTe substrate. These quantum dots permit to optically probe and address any spin state of a Mn atom in a controlled strain environment. The absence of strain induced magnetic anisotropy prevents an optical pumping of the Mn spin at zero magnetic field. Thus, a large photoluminescence is obtained under resonant optical excitation of the exciton-Mn complex. An efficient optical pumping of the coupled electronic and nuclear spins of the Mn is restored under a weak magnetic field. The observed reduction of the resonant photoluminescence intensity under magnetic field is well described by a model including the hyperfine coupling and a residual crystal field splitting of the Mn atom. Finally, we show that the second order correlation function of the resonant photoluminescence presents a large photon bunching at short delay which is a probe of the dynamics of coupled electronic and nuclear spins of the Mn atom.

\end{abstract}

\maketitle

\section{Introduction.}

The ability to control individual spins in semiconductors nanostructures is an important issue for spintronics and quantum information processing. The control of single spins in solids is a key but challenging step for any spin-based solid-state quantum-computing device \cite{Solotronics}. Thanks to their expected long coherence time, localized spins on magnetic atoms in a semiconductor host could be an interesting media to store quantum information in the solid state. Optical probing and control of the spin of individual or pairs of Mn atoms (S=5/2) have been obtained in self-assembled II-VI \cite{Besombes2004,Besombes2012} and III-V \cite{Kudelski2007,Krebs2013} semiconductor quantum dots (QDs). Recent studies of the spin dynamics of Mn atoms in self-assembled CdTe/ZnTe QDs have shown that a strain induced magnetic anisotropy ({\it i.e.} crystal field splitting), changing from dot to dot, blocks the electronic Mn spin along the QD growth direction \cite{LeGall2009,Gorycar2009,LeGall2010}: the precession of the Mn spin in the hyperfine field of its nuclei (I=5/2) or in a weak transverse magnetic field is quenched. This is at the origin of the Mn spin memory observed at zero magnetic field in II-VI QDs \cite{LeGall2009}.

As one aims at performing a fast optical coherent control of the spin of a magnetic atom, using for example the optical Stark effect \cite{Reiter2012} in a weak transverse magnetic field, the strain induced magnetic anisotropy of the Mn should be suppressed. Strain free Mn-doped QDs would be an interesting model system to probe the coherent spin dynamics of the magnetic atom in a weak magnetic field and would offer the possibility to externally tune the electron nuclei flip-flops with an applied magnetic field to prepare the nuclear spin of the Mn atom.

To obtain strain free magnetic QDs, we developed thin unstrained CdTe/CdMgTe quantum wells (QWs) doped with a low density of Mn atoms and lattice matched on a CdTe substrate. In these structures, localization of the carriers in the quantum well plane is achieved thanks to thickness fluctuations of the quantum wells at the monolayer scale \cite{Besombes2000}. A low density of Mn atoms is introduced during the quantum well growth and interface fluctuation islands containing individual Mn spins are obtained and optically probed using micro-spectroscopy techniques.

A large exciton-Mn (X-Mn) exchange interaction is obtained and a valence band mixing induced by the elongated shape of the confinement potential is observed both in the emission of neutral and charged magnetic QDs. The large exchange induced splitting of the exciton leads to an efficient thermalization of the X-Mn complex. To analyze the spin relaxation channels within the X-Mn complex in this new magnetic QD system, we performed resonant photoluminescence (PL) under excitation on the s-shell of the Mn-doped QDs. These measurements reveal a fast exciton spin relaxation and some slower spin-flips of the Mn during the lifetime of the exciton. In these strain free structures, we show that there is no optical pumping of the Mn spin at zero magnetic field. The optical pumping is restored under a weak magnetic field in the Faraday configuration. The magnetic field dependence of the resonant PL intensity is well described by a model including the Mn hyperfine coupling and a residual crystal field splitting likely due to alloy fluctuations around the Mn atom. The dynamics of the coupled electronic and nuclear spins of the Mn atom produces a large photon bunching revealed by the second order correlation function of the resonant PL signal.

The rest of the paper is organized as follows: In Sec. II, we describe the sample structure, the experimental techniques used and the basic optical properties of strain free singly Mn-doped QDs. In Sec. III, we analyze the influence of the valence band mixing on the spin properties of neutral and charged strain free Mn-doped QDs. In Sec. IV, we discuss the spin relaxation channels in the exciton-Mn complex. In Sec. V, we show that the optical pumping of the coupled electronic and nuclear spins of the Mn is controlled by a weak magnetic field applied along the QD growth axis. Finally, in Sec. VI, we present auto-correlation of the resonant PL of X-Mn that reveals the dynamics of the coupled electronic and nuclear spins of the Mn atom.

\section{Samples and experimental techniques.}

The studied sample consists of a 4 monolayers thick (1.6 nm) $CdTe/Cd_{0.7}Mg_{0.3}Te$ QW lattice matched to a (001) CdTe substrate. A low density of Mn atoms is introduced during the growth of the thin CdTe QW that is realized by Atomic Layer Epitaxy (ALE) at 280$^\circ$C. Both $CdTe/Cd_{0.7}Mg_{0.3}Te$ barriers (40nm thick below the QW and 90nm above the QW) are grown by molecular beam epitaxy. The ALE growth allows for a smoothing of the surface which leads to the presence of terraces with monomolecular steps \cite{Martrou1999} that are the source of in-plane localisation of the carriers. Non-magnetic and magnetic QDs containing a small number of Mn atoms (1,2,3 ...) are then formed.

\begin{figure}[hbt]
\includegraphics[width=3.5in]{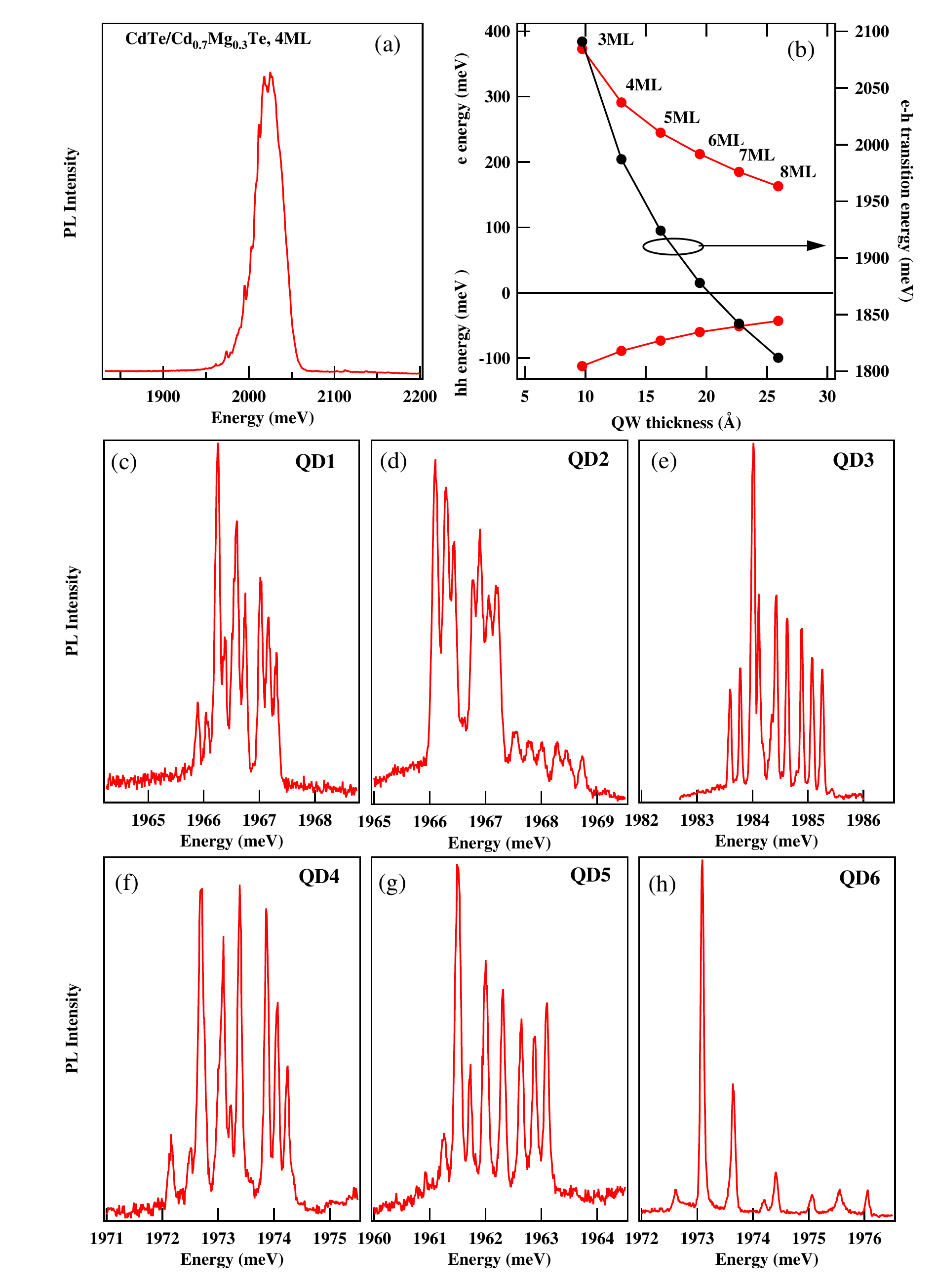}
\caption{(a) PL of a 4 monolayer CdTe/Cd$_{0.7}$Mg$_{0.7}$Te QW grown on a CdTe substrate. (b) Calculated energy levels of heavy hole (hh), electron (e) and electron-hole transitions in thin CdTe/Cd$_{0.7}$Mg$_{0.7}$Te QW. (c)-(h) PL of singly Mn-doped strain free QDs with increasing carrier-Mn overlap from QD1 to QD6 observed on the low energy side of the PL of a 4 monolayer CdTe/Cd$_{0.7}$Mg$_{0.7}$Te QW. The emission of QD2 is dominated by the contribution of a charged exciton, for the other QDs the neutral exciton is presented.} \label{Fig1}
\end{figure}

Optical addressing of individual QDs is achieved using micro-spectroscopy techniques. A high refractive index hemispherical solid immersion lens is mounted on the surface of the sample to enhance the spatial resolution and the collection efficiency of single dot emission in a low-temperature ($T$=5K) scanning optical microscope or in an He bath cryostat equiped with superconducting coils for magneto-optics measurements in Faraday geometry. Individual QDs are excited with a tunable continuous wave laser tuned to an excited state of the dots or on resonance with the QD s-state \cite{Glazov2007}. The resulting collected PL is dispersed and filtered by a 1 $m$ double monochromator before being detected by a cooled back-illuminated Si charge-coupled device camera. For measurements of the second order correlation function of the resonant PL, the circularly polarized collected photons are sent in a Hanbury Brown and Twiss (HBT) setup with a time resolution of about 0.7 ns. Under our experimental conditions with counts rates of a few kHz, the photon pair distribution measured with the HBT setup yields the intensity second order correlation function g$^{(2)}$($\tau$).

The PL of a 4 monolayers thick Mn-doped $CdTe/Cd_{0.7}Mg_{0.3}Te$ QW is presented in Fig.~\ref{Fig1}(a). The thickness fluctuations, and possible inter-diffusion of Cd and Mg, lead to the inhomogeneous broadening of the PL line of the QDs ensemble. The emission of non-magnetic and magnetic individual QDs is spectrally isolated on the low energy side of the broad PL line. The emission of six different strain free Mn-doped QDs observed on such structure are presented in Fig.~\ref{Fig1}(c)-(h) for an increasing value of the exciton-Mn exchange interaction. A X-Mn overall splitting as large as 3 meV can be obtained in some QDs (Fig.1(h)). This value of splitting is similar to the maximum splitting that one can observe in magnetic self-assembled CdTe/ZnTe QDs suggesting that in these two QDs systems the carrier-Mn overlap is mainly controlled by the confinement along the QD growth axis. In most of strain free Mn-doped QDs, the bright exciton presents a linearly polarized structure and a significant PL of the dark excitons is usually observed on the low energy side of the neutral exciton emission \cite{Leger2007}. Both characteristics are a consequence of the presence of valence band mixing in a low symmetry confinement potential \cite{Leger2007}.

\section{Carrier-Mn coupling in strain free Mn-doped quantum dots.}

To extract the value of the carrier-Mn exchange interaction and analyze the presence of valence band mixing in these strain free QDs, the experimental PL spectrum are compared with the exciton-Mn energy levels obtained from the diagonalization of a spin effective Hamiltonian similar to the one used to model the emission of Mn-doped self-assembled QDs \cite{Besombes2012,Fernandez2006,Glazov2007,Leger2007,Cao2011}. The spin interacting part of the Hamiltonian of a Mn-doped QD describing the coupling of the electron spin $\vec{\sigma}$, the hole spin $\vec{J}$, and a Mn spin $\vec{S}$ reads:
\begin{eqnarray}
{\cal H}_{XMn}=I_{eMn}\vec{\sigma}\cdot \vec{S} + I_{hMn}\vec{J}\cdot\vec{S} + I_{eh}\vec{\sigma}\cdot\vec{J} + {\cal H}_{scat}
\label{Hamilt1}
\end{eqnarray}
\noindent where the hole spin operators, $\vec{J}$ represented in the
basis of the two low energy heavy-hole states, are related
to the Pauli matrices $\tau$ by $J_z= \frac{3}{2}\tau_z$
and $J_{\pm}= \xi \tau_{\pm}$ with $\xi=-2\sqrt{3}e^{-2i\theta}\rho/\Delta_{lh}$. $\rho$ is the coupling energy between heavy-holes and light-holes split by the energy
$\Delta_{lh}$ and $\theta$ the angle relative to the [110]
axis of the principal axis of the anisotropy responsible
for the valence band mixing \cite{Leger2007}. I$_{hMn}$
(I$_{eMn}$) is the exchange integral of the hole (electron)
with the Mn atom. It strongly depends on the position of the Mn atom within the carrier wave function. I$_{eh}$ is the electron-hole exchange interaction which splits the bright and dark excitons and mixes the two bright excitons in the presence of valence band mixing in an anisotropic potential. The last term, ${\cal H}_{scat}$, describes the perturbation of the exciton wave function by the hole-Mn exchange interaction \cite{Besombes2005,Trojnar2013}. This perturbation depends on the value of the Mn spin S$_z$ and it can be represented, using perturbation theory, by an effective spin Hamiltonian ${\cal H}_{scat}=-\eta S_z^2$ with $\eta>0$. The perturbation of the wave function is responsible for the irregular energy spacing of the X-Mn and X$_2$-Mn lines observed in self-assembled Mn-doped QDs \cite{Besombes2005,Trojnar2013}.

\begin{figure}[hbt]
\includegraphics[width=3.5in]{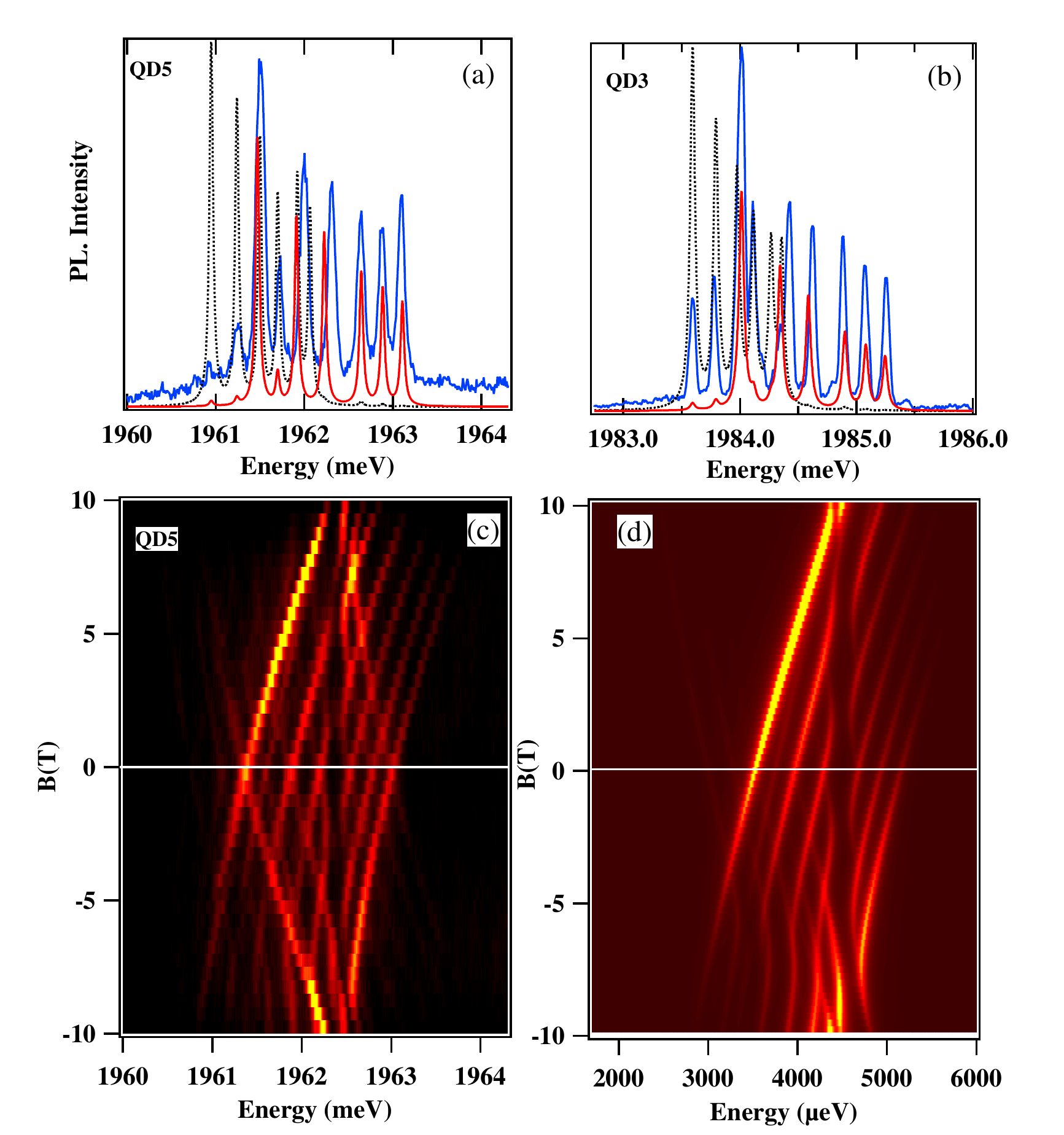}
\caption{Experimental (blue) and calculated (red: bright exciton, black: dark exciton) spectra of QD3 (a) and QD5 (b) at zero magnetic field. Intensity map of experimental (c) and calculated (d) magnetic field dependence of the PL of QD5. Exchange parameters used in the model are listed in table 1. An effective temperature T$_{eff}$=10K is used to reproduce the PL intensity distribution. For the magnetic field dependence of QD5 we use g$_e$=-0.55, g$_h$=0.75 and a diamagnetic shift coefficient $\gamma_{dia}$=1.25$\mu eVT^{-2}$.} \label{Fig2}
\end{figure}

A comparison between the calculated X-Mn spectra and the experimental PL data is given in figure~\ref{Fig2} for two QDs (QD3 and QD5). The main features of the experimental spectra can be well reproduced by this model. A determination of the hole-Mn ($I_{hMn}$), electron-Mn ($I_{eMn}$) and electron-hole ($I_{eh}$) exchange interaction energies can be obtained thanks to the observation of both bright and dark excitons levels at zero magnetic field (Fig.\ref{Fig2}(a) and Fig.\ref{Fig2}(b)). The values of these exchange parameters can be confirmed by the PL spectra obtained under a magnetic applied along the QD growth direction. The experimental magnetic field dependence of the PL of QD5 and the corresponding model are presented in Fig.\ref{Fig2}(c) and Fig.\ref{Fig2}(d) respectively. The exchange parameters, scattering parameter and valence band mixing amplitude estimated for the six strain free singly Mn-doped QDs discussed in this paper are listed in table \ref{table1}.

\begin{table}[htb] \centering
\caption{Values of the parameters used for the modelling of the six QDs discussed in this paper (I$_{eMn}$, I$_{hMn}$, I$_{eh}$ and $\eta$ are expressed in $\mu eV$). For all the QDs, we use $\theta$=0.}
\label{table1}\renewcommand{\arraystretch}{1.0}
\begin{tabular}{p{1.0cm}p{1.0cm}p{1.0cm}p{1.0cm}p{1.0cm}p{1.0cm}p{1.0cm}}
\hline\hline
&  QD1 & QD2 & QD3 & QD4 & QD5 & QD6\\
\hline
I$_{eMn}$ & $-70$ & $-55$ & $-90$ & $-110$ & $-100$ & $-190$\\
I$_{hMn}$ & $115$ & $135$ & $130$ & $170$ & $180$  & $330$\\
I$_{eh}$  & $-385$ & $ $ & $-430$ & $-520$ & $-515$ & $-550$\\
$\eta$  & $10$ & $10$ & $15$ & $25$ & $20$ & $20$\\
$\rho/\Delta_{lh}$ & $0.17$ & $0.1$ & $0.1$ & $0.1$ & $0.1$ & $0.05$\\
\hline\hline
\end{tabular}
\end{table}

To explain the dark exciton emission in the X-Mn spectra, a small valence band mixing term has to be introduced in the model. In strain free QDs, the mixing can only be induced by a shape anisotropy of the confinement potential. For an elongated QD, a hole band mixing appears through the non diagonal term of the Kohn-Luttinger Hamiltonian \cite{Bockelmann1992,Liao2012}. In the presence of such valence band mixing, the h-Mn exchange interaction couples bright and dark excitons and some PL from the dark excitons can be observed \cite{Leger2007}. In addition, the e-h exchange interaction can now couple bright exciton states and induce some linear polarization \cite{Leger2007}.

\begin{figure}[hbt]
\includegraphics[width=3.5in]{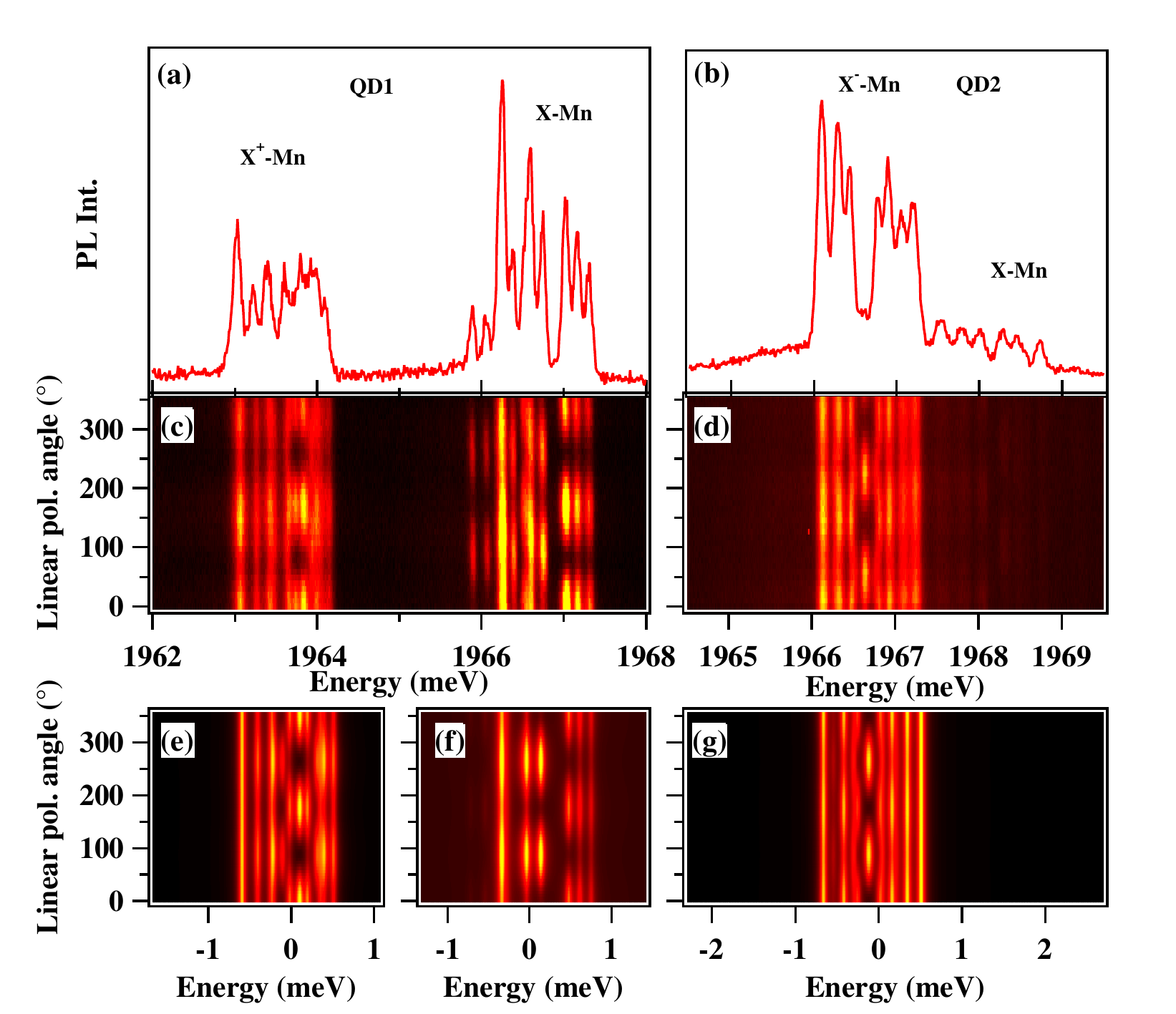}
\caption{PL of QD1 (a) and QD2 (b) where the neutral (X-Mn), the positively (X$^+$-Mn) and negatively (X$^-$-Mn) charged excitons coupled to a Mn spin can be observed. The corresponding intensity maps of the linear polarization dependence of the PL of the different excitonic species are shown in (c) and (d). They are compared to the results of the spin effective model in (e), (f) and (g). Parameters used in the model are listed in table~\ref{table1}}.
\ \label{Fig3}
\end{figure}

As shown in figure~\ref{Fig3}, the linear polarization induced by the valence band mixing can be observed in the PL of the neutral and charged Mn-doped QDs. The central lines of the charged excitons coupled to the Mn are linearly polarized. This can be explained by the spin flip interaction between the Mn and the hole induced by the valence band mixing \cite{Leger2006}. As presented in the bottom panel of Fig.~\ref{Fig3}, the main characteristics (number of emission lines, intensity distribution and linear polarization structure) of the linearly polarized emission spectra can be well reproduced by the spin effective model with a small valence band mixing coefficient and scattering parameter (see table~\ref{table1} for the parameters used in the model of QD1 and QD2). This confirms the validity of this spin effective model to describe strain free QDs doped with single Mn atoms.

\begin{figure}[hbt]
\includegraphics[width=3.5in]{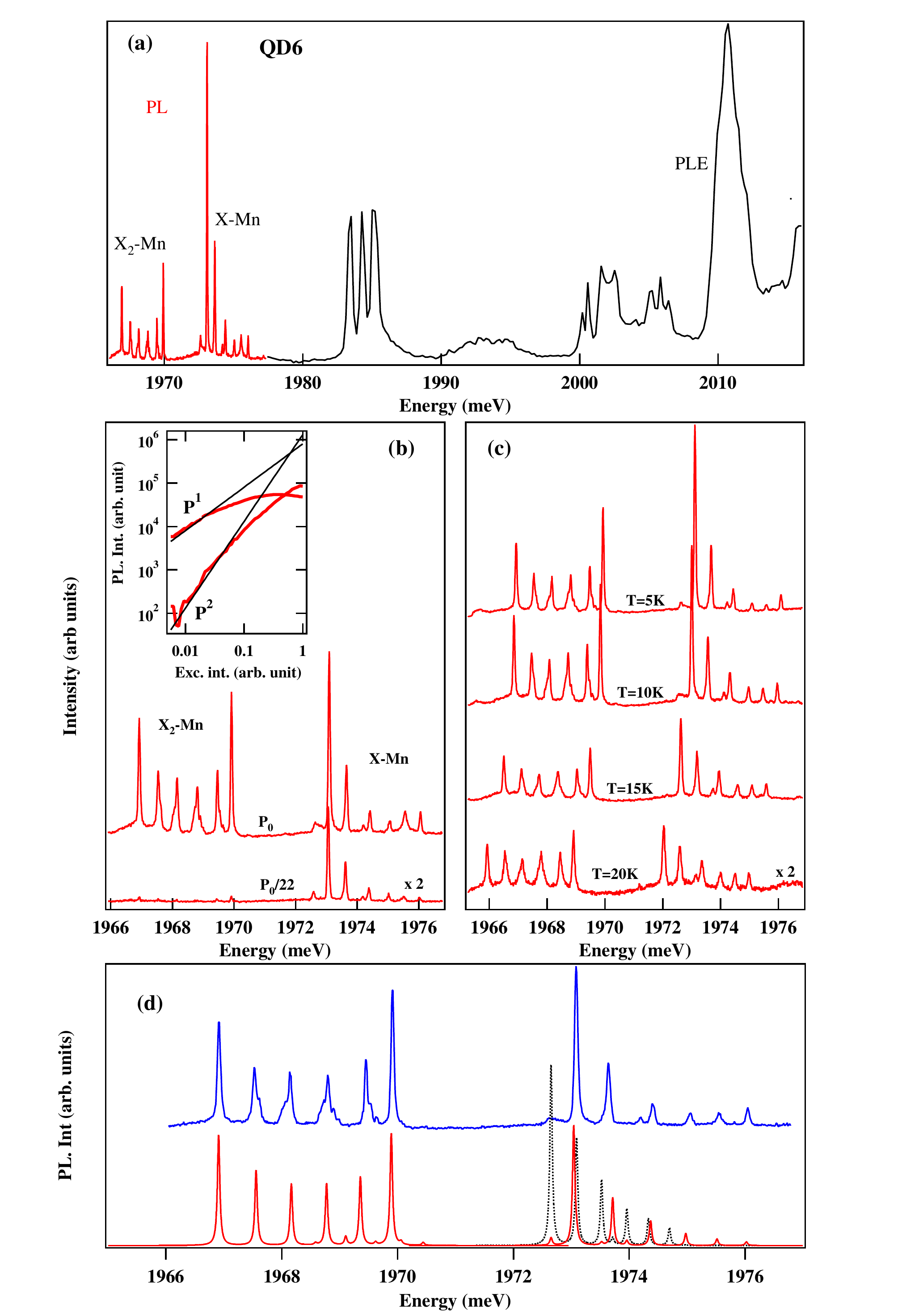}
\caption{(a) PL of X-Mn and X$_2$-Mn in QD6 and PL-excitation (PLE) detected on the low energy line of X-Mn. (b) Power dependence of the PL of X-Mn and X$_2$-Mn in QD6. (c) Temperature dependence of the PL of X-Mn and X$_2$-Mn in QD6. (d) Model of the emission of X-Mn and X$_2$-Mn including a thermalization with T$_{eff}$=5K (red: bright exciton and biexciton, black: dark exciton) compared with the experimental spectra (bleu).} \label{Fig4}
\end{figure}

The impact of ${\cal H}_{scat}$ can be emphasized in QDs with a large carrier-Mn overlap as for example QD6 in Fig.~\ref{Fig4}. For this QD, a large asymmetry appears in the emission intensity distribution at zero magnetic field. This asymmetry decreases with the increase of the lattice temperature (Fig.~\ref{Fig4}(c)). This is a signature of an efficient thermalization on the X-Mn split states. The significant transfer of population towards the low energy X-Mn levels results from efficient spin-flips within the X-Mn system during the lifetime of the exciton.

This thermalisation process also affects the biexciton-Mn emission (Fig.~\ref{Fig4}(b)). The perturbation term ${\cal H}_{scat}$, used in the model of neutral and charged excitons, induces a splitting of the biexciton-Mn states according to S$_z^2$. The combination of a large carrier-Mn exchange interaction and of a small energy splitting between the $s$-state and the first excited states (see figure~\ref{Fig4}(a)) in interface fluctuation QDs leads to a large perturbation of the excitonic wave functions by the hole-Mn exchange interaction. In the case of QD6, a perturbation term $\eta$=20$\mu$eV is used to model the irregular energy spacing of the X-Mn PL lines (Fig.~\ref{Fig4}(d)). This perturbation is two times larger for the biexciton and induces an overall X$_2$-Mn splitting of 240 $\mu eV$ (energy splitting between S$_z=\pm5/2$ and S$_z=\pm1/2$). The carrier relaxation mechanisms responsible for the thermalization of X-Mn also takes place within X$_2$-Mn. The thermalization produces a decrease of the intensity of the central lines of X$_2$-Mn ({\it i.e.} corresponding to S$_z$=$\pm$1/2). These intensity distributions in the PL of X-Mn and X$_2$-Mn can be well reproduced by the spin effective model (Fig.~\ref{Fig4}(d)) with an effective Mn spin temperature T$_{eff}$=5K.

The analysis of the optical emission of these strain free magnetic QDs demonstrates that they can be used at low temperature (typically below 20K, see Fig.~\ref{Fig4}(c)) to optically probe and address any spin state of a Mn atom in a strain free environment. We will show in the following how we can use such QDs to optically pump the coupled electronic and nuclear spins of a Mn atom.

\section{Exciton-Mn spin relaxation.}

An significant exciton relaxation is observed within the X-Mn complex under non-resonant excitation. More detailed information about the spin-relaxation mechanisms of the exciton exchange coupled with a Mn spin can be extracted from the energy and intensity of the PL signal obtained under resonant excitation on the ground state of the X-Mn complex. The spectral distribution of the PL under resonant excitation results from spin-flips processes within the 24 X-Mn levels and a simple thermalization should give rise to a thermal distribution on the intensities of the resonant PL spectrum.

Here, we report resonant PL results for two QDs presenting a very different carrier-Mn overlap (QD3 and QD6). In contrast to strained CdTe/ZnTe Mn-doped self-assembled QDs, a large PL signal is observed under resonant excitation on X-Mn in these strain free QDs. We will see in the next section that this results from the absence of optical pumping of the Mn spin at zero magnetic field. This is particularly interesting for a study of the X-Mn dynamics: the optical transition that is resonantly excitated is always absorbent and the energy of the re-emitted photons reveals the fastest spin relaxation channels.

\begin{figure}[t]
\includegraphics[width=3.5in]{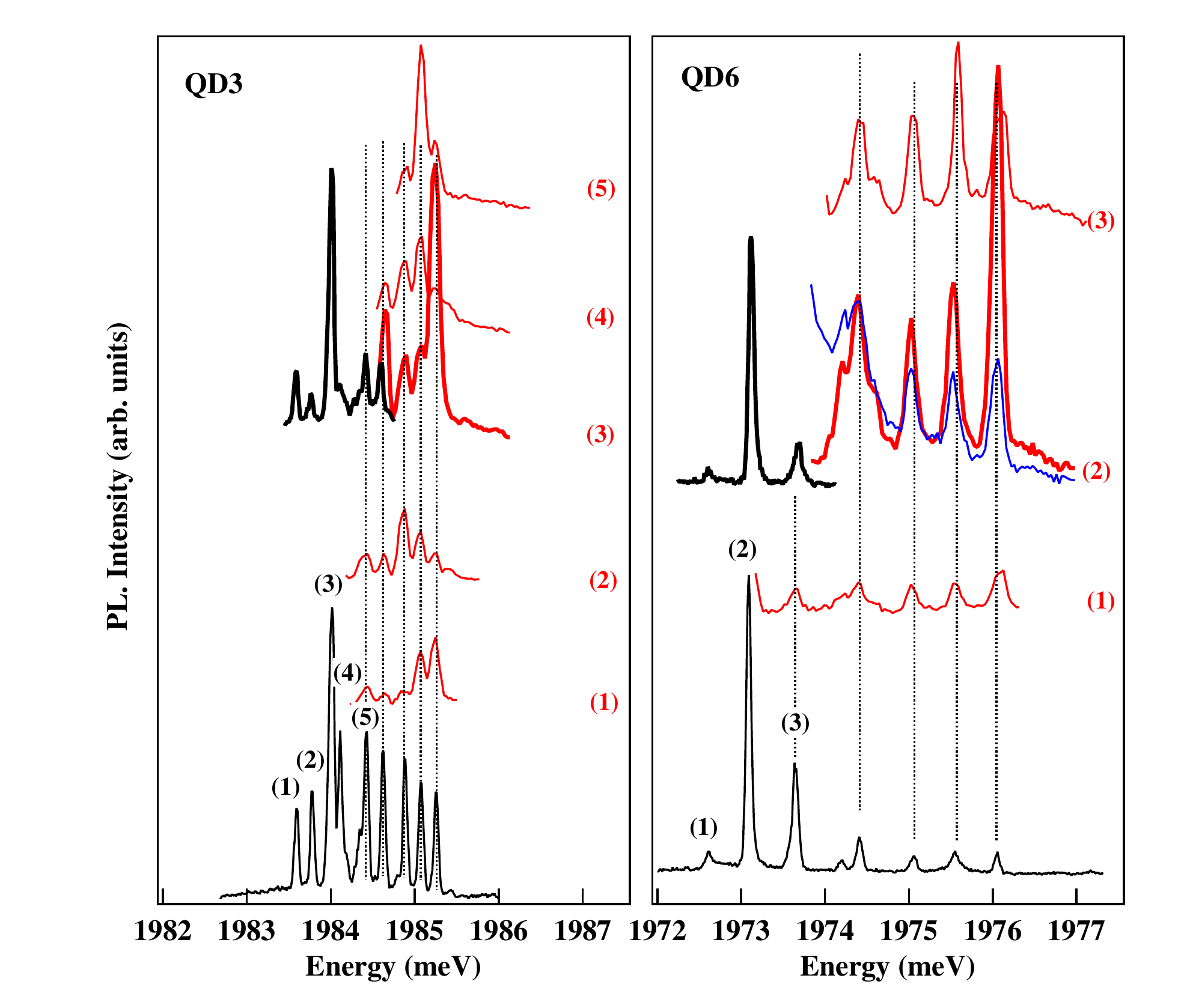}
\caption{PL (black), PLE (red) and resonant PL (bold black) spectra obtained on two QDs with very different X-Mn coupling, QD3 and QD6. The PLE is detected in cross-circular polarization configuration on the low energy lines (from (1) to (5)) as the energy of the laser is scanned on the high energy side of the X-Mn spectrum. The blue curve for QD6 is obtained with co-circular excitation-detection. The resonant PL (bold black) is obtained under circularly polarized excitation on the high energy line.} \label{Fig5}
\end{figure}

Figure~\ref{Fig5} presents the PL excitation spectra (PLE) detected on the low energy lines of X-Mn while the excitation laser is scanned on the high energy side of the spectrum. Corresponding resonant PL spectra obtained under excitation on the high energy line are also displayed and compared with non-resonant PL spectrum presented at the bottom of each panel. From these PLE and resonant PL spectra, it follows that the most efficient spin-relaxation channels within the X-Mn system corresponds to a spin-flip of the exciton with a conservation of the Mn spin. For instance, a resonant excitation on the high energy line in $\sigma+$ polarisation (excitation of S$_z$=+5/2) gives a maximum of emission in the low energy line in opposite circular polarization (detection of S$_z$=+5/2). In this spin-flip process, the exchange interaction with the magnetic atom can be considered in first approximation as a source of an effective quasi-static magnetic field giving rise to a splitting of the exciton. The relaxation of the bright excitons is then dominated by the interaction with acoustic phonons. Like in a large magnetic field, this spin relaxation process is enhanced by the large exciton splitting induced by the exchange interaction with the Mn spin \cite{Tsitsihvili2003,khaetskii2001,woods2002}.

Less efficient but still significant excitation transfers between levels corresponding to different spin states of the Mn are also observed especially in QDs with a large X-Mn splitting (QD6 in Fig.~\ref{Fig5}). These transfers under resonant excitation corresponds to the spin-flip channels required for a resonant optical pumping of the Mn spin.

\section{Resonant optical pumping of a Mn spin in a strain free quantum dot.}

\begin{figure}[hbt]
\includegraphics[width=3.5in]{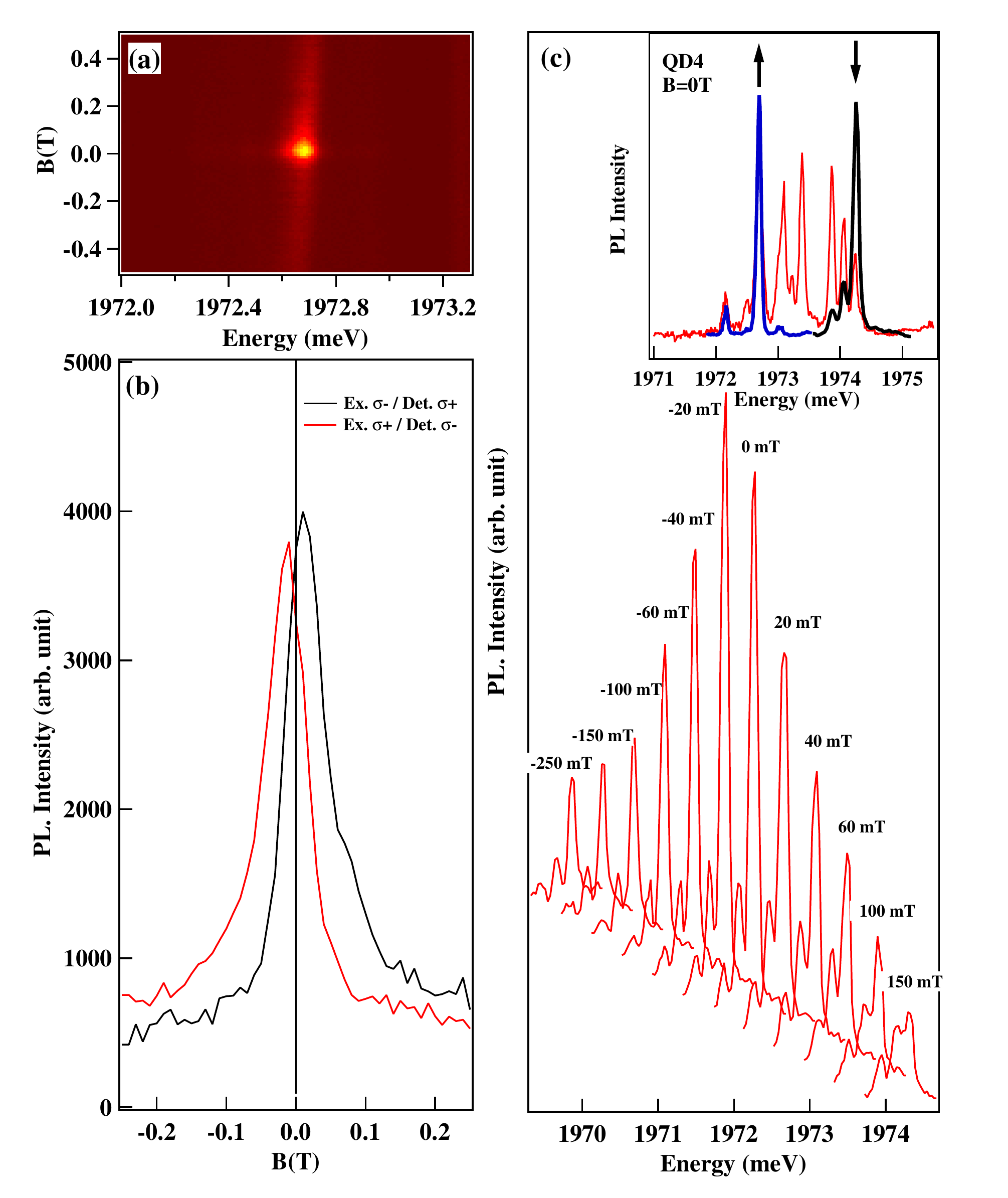}
\caption{Longitudinal magnetic field dependence of the intensity of the resonant PL in QD4 for crossed circular excitation and detection on the high and low energy lines respectively. (a) Intensity map of the magnetic field dependence of the resonant PL. (b) Magnetic field dependence of the resonant PL intensity for $\sigma-$ excitation / $\sigma+$ detection (black) and $\sigma+$ excitation / $\sigma-$ detection (red). (c) PLE detected on the low energy line for a $\sigma+$ excitation / $\sigma-$ detection and different transverse magnetic fields (the curves are horizontally and vertically shifted for clarity). The inset presents the non-resonant PL (red), the resonant PL for an excitation on the high energy line (blue) and the PLE detected on the low energy line (black) of QD4 at B=0T.} \label{Fig6}
\end{figure}

The very weak value of the crystal field splitting of the Mn expected in these strain free QDs is revealed by the longitudinal magnetic field dependence of the resonant PL intensity. As presented in figure~\ref{Fig6}, for a cross circularly polarized excitation and detection on the high and low energy lines of X-Mn (see inset of Fig.~\ref{Fig6}(c) for the excitation/detection configuration), a significant intensity of resonant PL is only observed for a longitudinal magnetic field lower than a few tens of mT. As a magnetic field is applied in the Faraday configuration, the resonant PL intensity abruptly decreases (Fig.~\ref{Fig6}(a)). A clear asymmetry is also observed in the magnetic field dependence of this resonant PL signal around zero Tesla. Under $\sigma-$ excitation on the high energy line and $\sigma+$ detection on the low energy line ({\it i.e.} excitation and detection of S$_z$=-5/2), the maximum of resonant PL intensity is slightly shifted towards positive magnetic fields and the decrease of the intensity is faster for negative magnetic fields (black curve in Fig.~\ref{Fig6}(b)). The situation is reversed for swapped excitation/detection circular polarizations (red curve in Fig.~\ref{Fig6}(b)).

This magnetic field dependence of the resonant PL is a consequence of the fine and hyperfine structure of the Mn atom on its optical pumping. To model this behaviour we consider the dynamics of the coupled electronic and nuclear spins of the Mn including a possible weak residual crystal field splitting. The fine and hyperfine Hamiltonian of a Mn atom in a CdTe layer grown along [001] axis is known from magnetic resonance measurements \cite{Qazzaz1995} and reads:

\begin{eqnarray}
\label{MnStrain} {\cal H}_{Mn}= {\cal
A}\overrightarrow{I}.\overrightarrow{S}
\nonumber\\
+\frac{1}{6}a[S_x^4+S_y^4+S_z^4-\frac{1}{5}S(S+1)(3S^2+3S-1)]
\nonumber\\
+{\cal D}_0[S_z^2-\frac{1}{3}S(S+1)]+E[S_x^2-S_y^2]
\nonumber\\
+g_{Mn}\mu_B\overrightarrow{B}.\overrightarrow{S}
\end{eqnarray}

\noindent where ${\cal A}$ is the hyperfine coupling (${\cal A}\approx+0.7\mu eV$) \cite{Causa1980} between the electronic (S=5/2) and nuclear (I=5/2) spins. The second term, $a=0.32\mu eV$ \cite{Causa1980} comes from the cubic symmetry of the crystal field and mixes different $S_z$ of the Mn spin. The deviation from the cubic symmetry due to strains in the QD or the presence of any other anisotropy at the Mn location are described by the additional crystal field terms ${\cal D}_{0}$ and $E$. ${\cal D}_{0}$ is the dominant term for a Mn atom in a highly strained self-assembled QD. It splits the Mn spin states according to S$_z^2$ and it is responsible for the long Mn spin life time at zero field \cite{LeGall2009}.

Different S$_z$ of the Mn are coupled by the non-diagonal terms of ${\cal H}_{Mn}$. For instance,
the hyperfine terms ${\cal A}$ couples two consecutive Mn spin states through an electron-nuclei flip-flop. An anisotropy of the Mn site also couples Mn spin states S$_z$ separated by two units through the crystal field term $E(S_x^2-S_y^2)$. In the absence of magnetic anisotropy (weak value of D$_0$), all these coupling terms at zero magnetic field prevent the optical pumping of the electronic Mn spin.

The dynamics of the coupled electronic and nuclear spins of the Mn is controlled by the time evolution of ${\cal H}_{Mn}$ and the electronic Mn spin relaxation time $\tau_{Mn}$. The nuclear spin relaxation is considered to be longer than the timescale of all the other spin relaxation mechanism discussed here. To describe the magnetic field dependence of the resonant PL, we extend here the model presented in reference \cite{Jamet2013}. The exciton states  $|J_z=-1,S_z=-5/2,I_z\rangle$ are laser coupled to the Mn states $|S_z=-5/2,I_z\rangle$. For a general description, valid from the low to the high optical excitation intensity regime, we consider that the resonant laser field induces a coherent coupling between the ground Mn spin state and the exciton state described by a Rabi energy $\hbar\Omega_{R}$. A pure dephasing time $\tau_d$ (\emph{i.e.} not related to an exchange of energy with a reservoir) is introduced for the exciton state. The X-Mn complex can relax its energy along a Mn spin conserving channel with a characteristic time $\tau_r$ (optical recombination of X) or along channels including a relaxation of the Mn spin with a spin flip time $\tau_{relax}$. The parameter $\tau_{relax}$ is a simplified effective way to describe the complex X-Mn spin dynamics at the origin of the optical pumping mechanism \cite{Cywinski2010,Cao2011}.

\begin{figure}[hbt]
\includegraphics[width=3.5in]{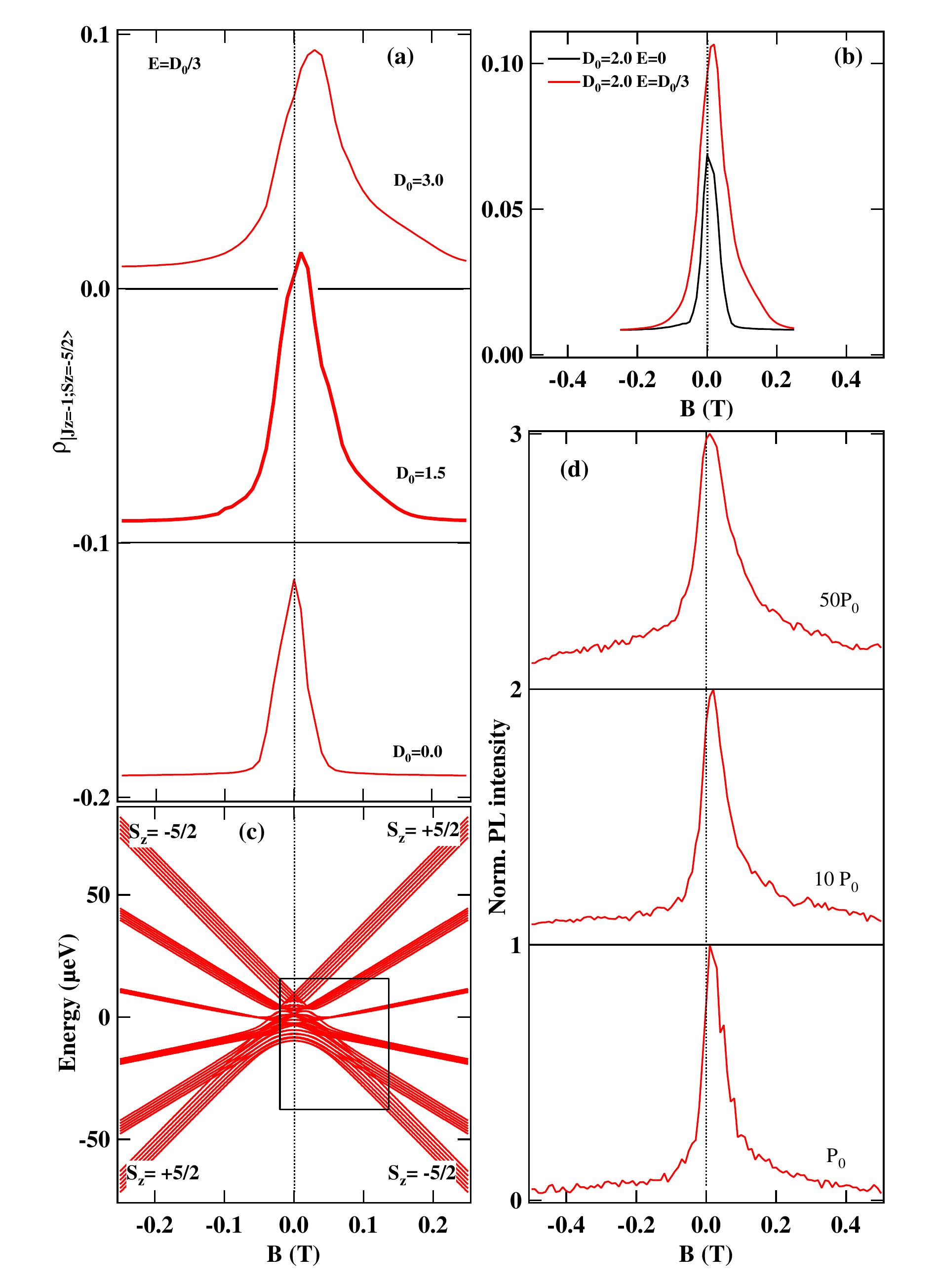}
\caption{(a) Modelling of the longitudinal magnetic field dependence of the population of the state $|J_z=-1,S_z=-5/2\rangle$ under resonant excitation for different values of the crystal field terms D$_0$ and E=D$_0$/3 (the curves are vertically shifted for clarity). The population of this level is proportional to the resonant PL signal. The parameters used in the calculation are $\hbar\Omega_R=15\mu eV$, $\tau_{Mn}=100ns$, $\tau_r=0.3 ns$, $\tau_{relax}=50 ns$, $\tau_d=100 ps$, $A=0.7\mu eV$, $a=0.32 \mu eV$ and $g_{Mn}=2$. (b) Calculated resonant PL signal for D$_0$=2$\mu eV$, E=0 (black) and E=D$_0$/3 (red). (c) Calculated fine and hyperfine structure of a Mn spin with $A=0.7\mu eV$, $a=0.32 \mu eV$, $g_{Mn}=2$, D$_0$=1.5$\mu eV$ and E=D$_0$/3. The black rectangle highlights the magnetic field range where the state S$_z$=-5/2 significantly interacts with the other Mn spin states. (d) Experimental magnetic field dependence of the resonant PL signal in QD4 for different excitation intensities and $\sigma-$ excitation / $\sigma+$ detection on the high and low energy lines respectively. The curves are normalized and vertically shifted for clarity.} \label{Fig7}
\end{figure}

With this level scheme, we can compute the steady state population \cite{Jamet2013} of the resonantly excited exciton level $\rho_{|J_z=-1,S_z=-5/2,I_z\rangle}$. This population is proportional to the intensity measured in the resonant PL experiment. The evolution under longitudinal magnetic field of the calculated population of the exciton states $|J_z=-1,S_z=-5/2,I_z\rangle$ is presented in Fig.~\ref{Fig7} for different values of the crystal field anisotropy D$_0$ and with E=D$_0$/3. We chose E=D$_0$/3 to obtain the maximum rhombic splitting and no preferential axial symmetry for the Mn environment\cite{Villain2006}. The main feature of the magnetic field dependence of the resonant PL observed experimentally ({\it i.e.} the width, the asymmetry and shift of the maximum from B=0T) can be well reproduced with D$_0$ in the 1-2$\mu eV$ range (Fig.~\ref{Fig7}(a)).

For a qualitative understanding of the behaviour of the resonant PL, we can look at the magnetic field dependence of the fine and hyperfine structure of the Mn atom (Fig.~\ref{Fig7}(c)). For a negative magnetic field, the state S$_z$=-5/2 is shifted away from the other Mn spin states: the electron-nuclei flip-flops are blocked, the optical pumping is restored within a few mT and the resonant PL signal decreases abruptly. For a positive magnetic field of a few mT, the state S$_z$=-5/2 is pushed to lower energy across the other Mn spin states. All the non-diagonal terms of ${\cal H}_{Mn}$ (in particular ${\cal
A}$ and $E$ terms) mixes the different S$_z$ and prevent the optical pumping of the electronic spin of the Mn. To restore the optical pumping, a larger positive magnetic field is needed. This explains that for a resonant excitation on S$_z$=-5/2, the drop of the resonant PL intensity is slower for positive magnetic fields. The magnetic field dependence is reversed for an optical excitation of S$_z$=+5/2 ({\it i.e} opposite circular polarization for the excitation). For large positive or negative magnetic fields, the Zeeman splitting of the Mn (controlled by g$_{Mn}$=2) dominates the hyperfine coupling and all the crystal field terms. The Mn spin is quantized along the direction of the applied magnetic field, the optical pumping efficiency increases and the resonant PL vanishes.

The Mn atoms are introduced in a thin CdTe layer that is coherently grown on a CdTe substrate. The first expectation would be that D$_0$=0. However, a residual crystal field in the $\mu eV$ range has to be introduced in the model to explain the experimental resonant PL data. For an axial crystal field ({\it i.e.} E=0), the calculated magnetic field dependence of the resonant PL is narrower and less asymmetric than in the experiment (Fig.~\ref{Fig7}(b)). This suggests that the remaining small crystal field felt by the Mn atom in our structure has no axial symmetry but arises from a more disordered environment. This crystal field could arise from an alloy composition fluctuation around the Mn atom. One should then keep in mind that the Mn atoms are located in a CdTe layer that is only 4 monolayers thick and surrounded by Cd$_{0.7}$Mg$_{0.3}$Te barriers. A Mn atom interacting with a confined exciton is always close to the barrier and the substitution of some Cd atoms by Mg atoms in the vicinity of the Mn reduces the symmetry of the local electric field and can induce weak crystal field terms D$_0$ and E \cite{Griscom1967, Wang2012}. While further investigations would be needed to fully understand this point, we observe that D$_0$ is positive as one could expect from the lattice parameter difference between MgTe and CdTe: MgTe has a smaller lattice parameter than CdTe, so the presence of a Mg atom in the vicinity of the Mn induces a local reduction of the lattice parameter and then a positive crystal field term \cite{Causa1980}.

As presented in Fig.~\ref{Fig7}(d), the magnetic field dependence of the resonant PL is observed in a broad range of excitation intensities. However, the width of the resonant PL peak and the residual background of resonant PL increase with the increase of the excitation intensity. This reflects a reduction of the optical pumping efficiency at high excitation power. This reduction at high excitation power is likely due to a perturbation of the Mn spin by the additional non-resonant injection of free carriers that decreases the Mn spin memory \cite{Besombes2008}.

\section{Dynamics of the resonant photoluminescence of X-Mn.}

The dynamics resulting from the coupling of the electronic and nuclear spins of the Mn in this weak crystal field can be observed in the second-order correlation function of the resonant PL signal of X-Mn. Such experiment in strain free Mn-doped QDs is made possible by the absence of optical pumping of the Mn at zero magnetic field and the resulting large resonant PL intensity.

We performed auto-correlation of the resonant PL using an HBT set-up \cite{Besombes2008} with a time resolution of about 0.7 ns. In these start-stop experiments, the detection of the first photon indicates by its energy and polarisation that the Mn spin has a given orientation. The detection probability of a second photon with the same energy and polarisation is then proportional to the probability of conserving this spin state. The time evolution of this intensity correlation signal is a probe of the spin dynamics of the Mn atom and of the X-Mn complex.

\begin{figure}[hbt]
\includegraphics[width=3.5in]{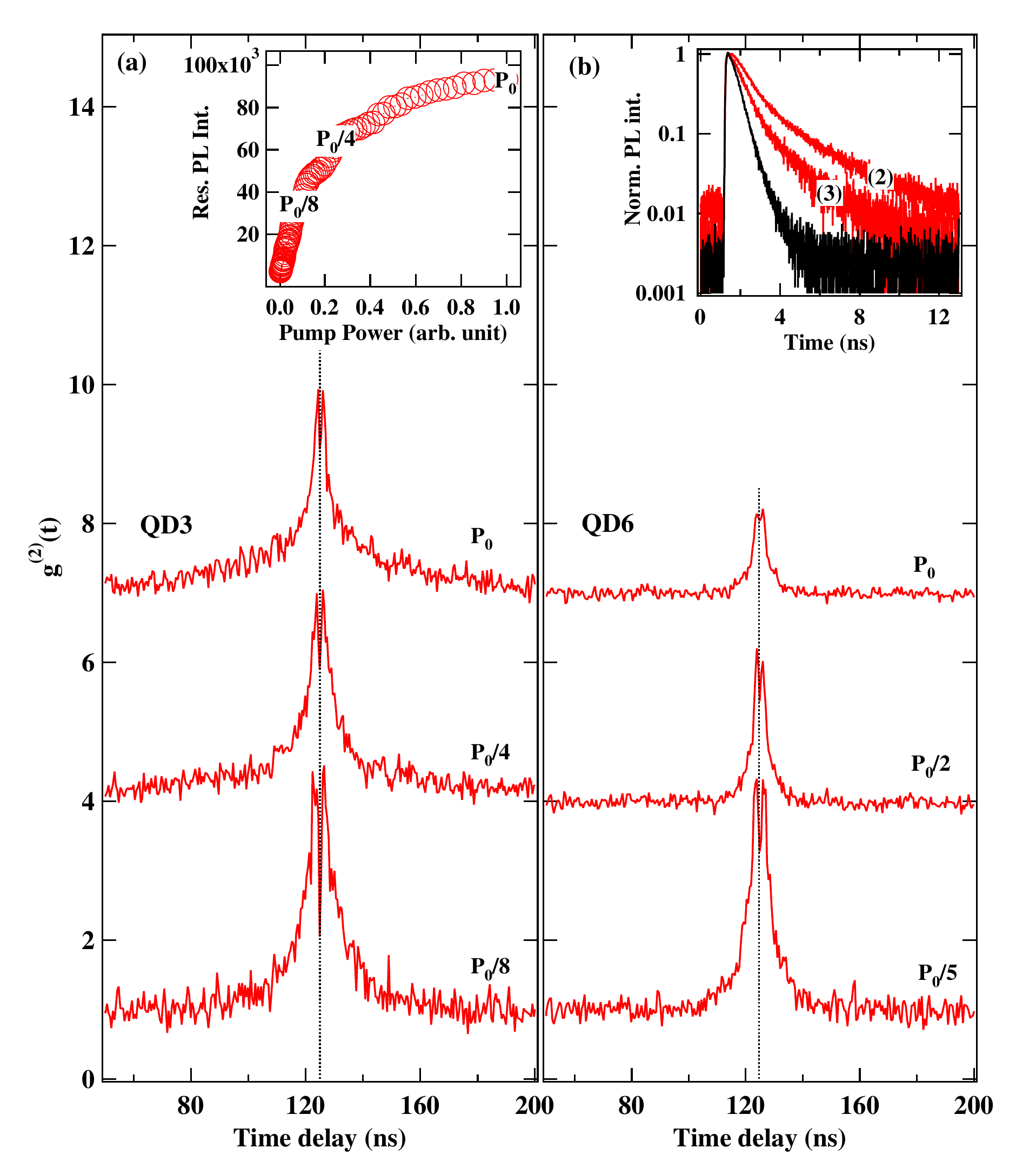}
\caption{Excitation power dependence of the auto-correlation of the resonant PL obtained on QD3 (a) and QD6 (b) for cross-circular excitation and detection on the high energy and low energy lines respectively. Inset in (a): excitation power dependence of the resonant PL intensity of QD3. Inset in (b): PL decay time of the two low energy lines of QD6 (line (2) and (3) in Fig.~\ref{Fig6}). The black curve corresponds to the PL decay of the biexciton in QD6 where no contribution of the dark exciton at long time delay is observed.} \label{Fig8}
\end{figure}

Examples of excitation power dependence of the auto-correlation of the resonant PL are given in figure~\ref{Fig8} for two QDs with a very different exciton-Mn overlap (QD3 and QD6). The data are obtained under cross-circular excitation an detection on the high and low energy lines of X-Mn respectively (experimental configuration illustrated in the inset of Fig.~\ref{Fig6}(c)). With this configuration, we resonantly excite and detect the same spin state of the Mn (either S$_z$=+5/2 for a $\sigma+$ excitation or S$_z$=-5/2 for a $\sigma-$ excitation).

\begin{figure}[hbt]
\includegraphics[width=3.5in]{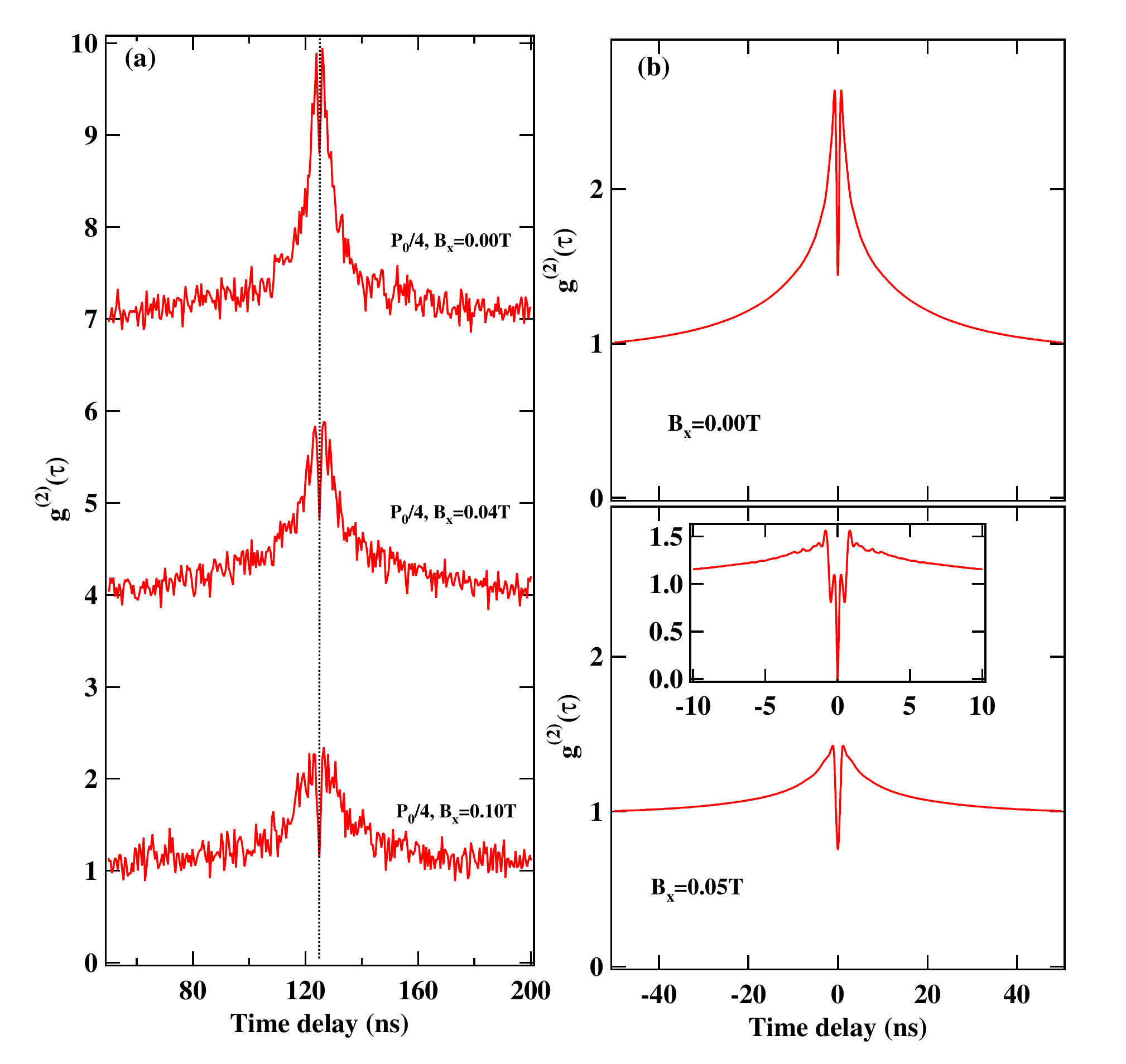}
\caption{(a) Transverse magnetic field dependence of the auto-correlation of the resonant PL of QD3 (The curves are vertically shifted for clarity). (b) Calculated auto-correlation signal with and without transverse magnetic field B$_x$. The curves are obtained after convolution with the response of the HBT setup ($\approx0.7ns$). The inset shows the calculated autocorrelation at short time delay before convolution with the response set-up. The parameters used in the calculation are: $1/g=0.3 ns$, $\tau_{Mn}=100ns$, $\tau_{XMn}=10ns$, $\tau_r=0.3 ns$, $\tau_{nr}=10 ns$ $\tau_{relax}=50 ns$, $A=0.7\mu eV$, $a=0.32 \mu eV$, $g_{Mn}=2$, D$_0$=1.5$\mu eV$, E=D$_0$/3 and the QD parameters of QD3 (see table \ref{table1}).} \label{Fig9}
\end{figure}

The auto-correlation signal presents first a dip near zero delay characteristic of a single-photon emitter. The width of this antibunching signal is given by the lifetime of the emitter (inset of Fig.~\ref{Fig8}(b)) and the generation rate of excitons. Its depth is limited here by the time resolution of the HBT setup. In addition, the large bunching at short delays reveals a significant PL intermittency. The width and the amplitude of the bunching decreases with the increase of the excitation intensity. The decrease of the amplitude of this bunching signal in a transverse magnetic field of a few tens of mT (Fig.~\ref{Fig9}(a)) confirms that the PL intermittency results mainly from fluctuations of the Mn spin.

To model this bunching signal, we have to take into account the spin dynamics of the Mn and X-Mn. In a simplified level scheme, we neglect the hyperfine splitting within the X-Mn complex and we consider that a single exciton state, $|J_z=-1\rangle$, is laser coupled to the Mn spin states $|S_z=-5/2,I_z\rangle$ with a generation rate $g$. The dynamics of the Mn spin in an empty QD is described by the time evolution of ${\cal H}_{Mn}$ and the relaxation time $\tau_{Mn}$ describing the coupling by spin-flips of one unit the different spin states $S_z$.

The 24 X-Mn levels are obtained from the diagonalization of ${\cal H}_{XMn}$. X-Mn can optically recombine along a Mn spin conserving channel with a lifetime $\tau_r$ for the bright exciton or $\tau_{nr}$ for the dark exciton. X-Mn can also relax along channels involving a change of the spin by one unit with a characteristic time $\tau_{XMn}$. This relaxation path, combined with an optical recombination of the exciton, permits a transfer of population from the exciton state $|J_z=-1,S_z=-5/2\rangle$ to any other spin state of the Mn $S_z$. To describe the main X-Mn spin relaxation channel revealed by the resonant PL spectra ({\it i.e} relaxation of the exciton with conservation of the Mn spin observed in Fig.\ref{Fig5}), we consider a transfer rate between bright excitons with a characteristic time $\tau_{X}$.

At finite temperature, the intraband relaxation rates $\Gamma_{\gamma\rightarrow\gamma'}$ between the different states of the X-Mn complex depend on their energy separation $E_{\gamma\gamma'}=E_{\gamma'}-E_{\gamma}$. Here we use $\Gamma_{\gamma\rightarrow\gamma'}$=1/$\tau_{i}$ if $E_{\gamma\gamma'}<0$ and $\Gamma_{\gamma\rightarrow\gamma'}$=1/$\tau_{i}e^{-E_{\gamma\gamma'}/k_BT}$ if $E_{\gamma\gamma'}>0$ ($\tau_i$ corresponding either to $\tau_{XMn}$ or $\tau_{X}$) \cite{Govorov2005}. This describes a partial thermalization among the 24 X-Mn levels during the lifetime of the bright or dark excitons, as observed experimentally (see Fig.~\ref{Fig4}).

In this model we also consider that for a magnetic field lower than a few hundreds mT, the Zeeman energy of the exciton can be neglected since it is much smaller than the X-Mn exchange interaction. We only take into account the effect of the magnetic field on the Mn in the empty QD (last term of ${\cal H}_{Mn}$).

Using this level scheme, we can calculate the time evolution of the 60x60 density matrix describing the population and the coherence of the 36 states of the Mn alone (empty QD described by ${\cal H}_{Mn}$) and the 24 X-Mn states described by ${\cal H}_{XMn}$. To model the auto-correlation of the resonant PL, the initial state of the system is set to $|S_z=-5/2,I_z\rangle$ and one monitors the time evolution of the population of $|J_z=+1,S_z=-5/2\rangle$ (low energy line) for an excitation of  $|J_z=-1,S_z=-5/2\rangle$ (high energy line) with a generation rate $g$. When normalized to one at a long time delay, this time evolution accounts for the auto-correlation function of the transition associated with the level $|J_z=+1,S_z=-5/2\rangle$.

Auto-correlation signals calculated with and without transverse magnetic field are presented in Fig.~\ref{Fig9}(b). In a weak transverse magnetic field (inset of Fig.~\ref{Fig9}(b)), the auto-correlation displays small oscillations at short delay due to the precession of the Mn electronic spin. The limited time resolution of the HBT setup prevents the observation of these oscillations. To take into account this time resolution, the calculated curves are convoluted with the response of the system (Fig.~\ref{Fig9}(b)). As in the experiment, the calculated auto-correlation presents a large bunching at short time delay with a typical half width at half maximum of about 4 ns. This short Mn spin memory is due to its precession in the hyperfine field of its nuclei \cite{Scalbert2013}. At low excitation intensity and at zero magnetic field, the width of the experimental photon bunching is well reproduced by the model.

In a weak transverse magnetic field, the amplitude of the bunching signal decreases significantly. The transverse magnetic field dependence of the shape of the calculated bunching signal (width and amplitude) is in qualitative agreement with the experiment (Fig.~\ref{Fig9}). A magnetic field of a few tens of $mT$ has no effect on the X-Mn system and do not change the contribution of the X-Mn dynamics on the bunching amplitude. The observed and calculated decrease of the bunching amplitude is then induced by the precession of the Mn spin in the transverse magnetic field when the QD is empty. This confirms that in our experiments the bunching signal at zero field is mainly controlled by the dynamics of the coupled electronic and nuclear spins of the Mn in a weak residual crystal field.

The width of the bunching signal measured at low excitation power is in good agreement with the calculated Mn spin dynamics. However, at high excitation intensity a reduction of the width and the amplitude of the bunching is observed (Fig.~\ref{Fig8}). This reduction cannot be explained by the presented model and it likely comes from a perturbation of the Mn spin by non-resonantly injected free carriers in the vicinity of the QD. A similar increase of the Mn spin relaxation rate was already observed for much confined CdTe/ZnTe QDs under non-resonant excitation \cite{Besombes2008}.

\section{Conclusion}

We have incorporated and optically probed individual Mn atoms in strain free CdTe QDs formed by interface fluctuations in thin quantum wells. Despite the weak lateral confinement and a possible valence band mixing induced by an anisotropic shape of the confinement potential in these natural QDs, we have shown that an individual Mn atom can be optically addressed by a resonant optical excitation. Magnetic field dependence and auto-correlation of the photoluminescence signal obtained under resonant excitation of the QD ground state show that, in these strain free structures, an isolated Mn atom experiences a residual weak crystal field splitting likely due to local alloy fluctuations. An efficient optical pumping the Mn spin is obtained under a weak magnetic field in the Faraday geometry.

Despite this residual crystal field, strain free magnetic QDs are promising systems to probe the coherent dynamics of the coupled electronic and nuclear spins of a Mn atom in zero or weak magnetic field. These strain free Mn-doped QDs could also be included in photonic structures such as optical planar micro-cavities to increase the number of collected photons and their interaction with light. This would open the possibility to realize an optical coherent control of an individual Mn spin in a weak transverse field using the optical Stark effect or to probe and control the dynamics of the nuclear spin of the Mn atom.

\begin{acknowledgements}
This work was partly supported by the European Initial Training Network Spinoptronics.
\end{acknowledgements}

\end{document}